# Tunnel Junction Enhanced Nanowire Ultraviolet Light Emitting Diodes

ATM Golam Sarwar[1], Brelon J May[2], and Roberto C Myers[1,2].

[1]Department of Electrical and Computer Engineering, The Ohio State University, Columbus, OH 43210.

[2]Department of Materials Science and Engineering, The Ohio State University, Columbus, OH 43210.

Abstract:

Polarization engineered interband tunnel junctions (TJs) are integrated in nanowire ultraviolet (UV) light emitting diodes (LEDs). A ~6V reduction in turn-on voltage is achieved by the integration of tunnel junction at the base of polarization doped nanowire UV LEDs. Moreover, efficient hole injection into the nanowire LEDs leads to suppressed efficiency droop in TJ integrated nanowire LEDs. The combination of both reduced bias voltage and increased hole injection increases the wall plug efficiency in these devices. More than 100 µW of UV emission at ~310 nm is measured with external quantum efficiency in the range of 4 - 6 m%. The realization of tunnel junction within the nanowire LEDs opens a pathway towards the monolithic integration of cascaded multi-junction nanowire LEDs on silicon.

Ultraviolet (UV) light has several important applications such as water and air purification, chemical agent detection, adhesive curing and nano-(micro-)scale photolithography[1]. AlGaN based light emitting diodes (LEDs) are the main source for solid state UV emitters mainly because of the wide range of direct band gaps spanning the UVA, UVB, and UVC range[2]–[6]. However, these LEDs suffer from several fundamental material problems, especially poor hole doping due to large acceptor ionization energy (630 meV in AlN and 200 meV in GaN) which causes low injection efficiency (due to low hole density) and high electrical losses (due to high p-type contact resistance).

An attractive alternative to the epilayer approach utilizes nanowires that allow for integration of highly lattice mismatched materials without creating dislocations [7], [8]. UV emission spanning from GaN (365 nm) to AlN (210 nm) band gap wavelengths was demonstrated using AlGaN nanowires in the past few years[9]–[12]. A wavelength tunable polarization doped nanowire heterostructure has been used to demonstrate the first UVC nanowire LEDs [9]. The same heterostructure design was later used to study the free hole density in polarization graded AlGaN nanowires[13]. This study reveals that p-type polarization doping in graded composition AlGaN nanowires is limited by deep donor states and free hole density comparable to bound polarization charge can be achieved only when extreme compositional grading is used.

Fig. 1 shows the schematic diagram of a wavelength tunable polarization doped nanowire UV LED structure[9], [10], [13]. Catalyst free GaN nanowires exhibit a dominant N-face polarity[14] which induces p-type (n-type) conductivity at the base (top) of the nanowire when compositionally graded from GaN to AlN (AlN to GaN). Wavelength tunable multiple AlGaN quantum disks are inserted at the center of the nanowire separated by AlN barriers to serve as the active region. To make a p-type electrical contact at the bottom of the nanowires these doubly graded heterostructures are grown on p-type Si or high work function metals[9], [13], [15]. As shown in the energy band diagram (Fig. 1(b)), when such a nanowire heterostructure is grown on a p-type Si wafer it results in a large valence band discontinuity ($\Delta E_V \sim 1.8$ eV) that increases the LED turn-on voltage. The ineffective hole injection from the p-type Si wafer into the p-type graded AlGaN region remains a key bottleneck for realizing high efficient nanowire UV emitters.

In this work, we integrate polarization-engineered tunnel junctions (TJs) [16]–[18] within the nanowire UV LEDs. The TJs efficiently inject holes into the p-type graded layer utilizing interband tunneling, thereby converting the bottom contact from p-type to n-type. As a result, the TJ integrated nanowire LEDs can be grown on n-type Si wafers, where, due to the small conduction band discontinuity ($\Delta E_C \sim 0.5$ eV) between Si and GaN, charge transport at the Si/GaN interface is greatly enhanced. Overall, this results in ~6V reduction in turn-on voltage of the nanowire UV LEDs. Furthermore, the efficiency droop in these nanowire LEDs is greatly reduced, which we attribute to enhanced non-equilibrium hole concentration due to interband tunneling.

Fig. 2(a) shows the schematic diagram of the TJ integrated nanowire UV LEDs. 100 nm Si doped GaN nanowires are grown on n-type Si wafer. A polarization engineered n++ GaN/ InGaN/ p++ GaN interband TJ [16]–[18] is deposited followed by a doubly graded polarization doped UV LED structure as described earlier [9], [10], [13]. The nanowire heterostructures are grown from self-assembled catalyst free nanowires using a two-step [19]–[21] dynamic growth process byplasma assisted molecular beam epitaxy in a Veeco GEN 930 system. The InGaN insertion in the TJ is grown at relatively low temperature (~600C) for effective In incorporation while the AlGaN active region is grown at relatively high temperature (~840C) to suppress optically active defects[9]. The AlGaN graded layers are doped with Mg (at the bottom) and with Si (at the top). The transition from low temperature TJ growth to high temperature graded AlGaN growth is found to be a very important factor. The optimization of this process along with the detailed growth method is described in supporting material. Fig 2(b) shows the resulting energy band diagram of TJ integrated nanowire UV LEDs. One advantage of the TJ can be easily understood by looking at the Si/GaN interface of the energy band diagram (compare Fig. 2(b) and Fig. 1(b)). The carrier injection barrier at the Si/GaN interface is greatly reduced compared to that shown in Fig. 1(b). Forward biasing the polarization doped UV LED structure causes the TJ to be at reverse-bias. The interband tunneling at the TJ injects electrons into the n-GaN base (which escape to the n-type Si contact) and holes into the p-type graded layer. These holes along with the electrons from the n-type graded layer diffuse into the AlGaN active region and go through radiative recombination to emit UV photons.

Fig. 2(c) shows tilted view scanning electron microscopy (SEM) image of as grown TJ integrated nanowire UV LEDs. The SEM image shows inverse tapering morphology which is attributed to enhanced radial growth during low temperature TJ growth due to reduced adatom mobility. The nanowire arrays exhibit nearly uniform height and diameter.

Room temperature photoluminescence (PL) spectrum of as grown nanowires is shown in Fig. 3. The nanowires are optically excited with a third harmonic (~266 nm) of a Ti:sapphire oscillator mode-locked at 800 nm. The sharp peak at 400 nm is from the 2$^{nd}$ harmonic of the oscillator. Apart from this, the PL spectrum shows three different peaks. The ~312 nm peak corresponds to optical recombination in the AlGaN active region. The 360 nm peak corresponds to the GaN nanowire base. The broad peak at the visible wavelengths centered at 500 nm could be a result of optical recombination in the InGaN insertion in the TJ or from deep level defects in the AlGaN region.

Nanowire LED devices are fabricated from the grown nanowire array. A 10 nm Ti/ 20 nm Au semitransparent contact is deposited as a top metal contact. In diffused contact is formed on n-type Si wafer after mechanical removal of the nanowires. Fig. 4 shows the current-voltage (I-V) characteristics of nanowire UV LEDs with (red) and without (blue) TJ integration. The inset shows I-V characteristics in log scale. Both I-Vs show good diode characteristics with very small reverse bias leakage. Moreover, the LED with the TJ shows less reverse bias leakage current compared to the LED without the TJ. The LED with the TJ also shows ~6V reduction in the turn-on voltage. This reduction in turn-on voltage is attributed to the reduction of the carrier injection barrier at the bottom Si contact.

We perform electroluminescence (EL) characterization on the nanowire UV LEDs. Fig. 5(a) and (b) shows EL spectra from nanowire LEDs without and with TJ under dc current injection, respectively. UV emission at ~ 310 nm is observed from both LEDs. This indicates that the incorporation of TJ at the base of the nanowire LED does not alter the emission characteristics of the AlGaN active region. Fig. 5(c) shows integrated EL vs current density under dc current injection. The nanowire LEDs with the TJ can undertake ~2× current compared to the LEDs without the TJ. This is the direct effect of turn-on voltage scaling in the LEDs with the TJ. The reduced operating voltage reduces the Joule heating in the LEDs with the TJ; as a result they can be operated at high injection current. Utilizing the high injection current at least ~3.5× more light (integrated EL) emission is detected from the LEDs with the TJ. Moreover, TJ integrated LEDs emit ~2× more light at 180A/cm$^2$, which is the maximum operating current of LEDs without the TJ.

Fig. 5(d) shows the relative external quantum efficiency (EQE) as a function of current density. Relative EQE is calculated from the ratio of integrated EL and injection current. It is observed that the nanowire LED without the TJ (blue circles) shows peak efficiency at very small current (~1-2 A/cm$^2$) and exhibits extreme efficiency droop with increased injection current (~66% @ 180A/cm$^2$). This phenomena is attributed to the poor hole injection at the nanowire/Si interface in these LEDs. The effect of poor hole injection is schematically shown in Fig. 1(b). Under forward bias, electrons are injected from the top contact (right in Fig. 1(b)) and holes are injected from the bottom nanowire/Si interface. With the increased forward bias, electrons can be injected efficiently while hole injection is inefficient due to the large injection barrier at the bottom Si/GaN interface and results in low hole concentration in the p-graded layer. As a result the emission in the active region is limited by poor hole injection. The injected excess electrons overflow the active region causing an electron drift current in the p-graded layer (because of the quasi electric field in the conduction band). The overflown electrons eventually recombine in p-type Si contact and resulted in an extreme efficiency droop (blue circles in Fig. 5(d)).

On the other hand integration of a TJ allows hole injection into the p-type graded layer through an interband tunneling process (see Fig. 2 (b)) without a hole injection barrier. Thus, a high concentration of holes is available in non-equilibrium in the p-type graded region. This suppresses electron overflow because availability of holes ensures the efficient capture of the injected electrons in the active region and results in a higher efficiency in the TJ devices. We observe a shift of the peak EQE toward high current in the TJ integrated LED devices. This indicates increased Shockley–Read–Hall recombination in the TJ integrated LEDs. However, we found that the peak EQE shift can be reduced as well as high peak EQE value can be obtained by optimizing the temperature transition from low temperature TJ growth to high temperature

AlGaN growth (see supporting information). Relatively small efficiency droop (18% @ 180A/cm$^2$ for optimized TJ LEDs – see supporting information) is observed in TJ LEDs.

The use of a TJ, on one hand increases EQE of the nanowire UV LEDs by efficient hole injection and on the other hand decreases electrical losses by decreasing the operating bias voltage. The combined effect can be observed when integrated EL (~optical output power) vs input power is plotted in Fig. 5(e). The ratio of integrated EL from LED with the TJ to LED without the TJ is also plotted in Fig. 5 (e) (right axis). The optical output from LED device with the TJ surpasses the optical output from the LED without the TJ at a very low injection power of 0.15 kW/cm$^2$. The ratio increases with increasing input power and reaches 2.73 at an input power of 3 kW/cm$^2$. Fig. 5(f) shows relative wall plug efficiency (WPE) of the LEDs with (red squares) and without (blue circles) the TJ. The WPE of the TJ LED reaches maximum at an input power of 1 kW/cm$^2$ at which point the LED with the TJ gives at least ~2× light output.

We also fabricate large area devices utilizing a spin on glass (SOG) based planarization technique, 5 nm Ti/ 10 nm Au semitransparent metal contact, and thick Ti/Al/Ni/Au metal grid. Fig. 6 shows on-wafer power measurement from a 1 mm$^2$ TJ integrated nanowire UV LED. More than 100 μW power is measured under DC excitation. The EQE is measured in the range of 4 – 6 m%.

In conclusion, we integrate polarization engineered tunnel junction with nanowire UV LEDs. The incorporation of a tunnel junction increases hole injection efficiency and suppresses the extreme efficiency droop in nanowire UV LEDs. The reduced operating voltage along with increased hole injection increases the light output from the LEDs integrated with the TJ and shows increased wall plug efficiency. The ability to integrate tunnel junctions in III-N nanowire

LEDs can lead to the monolithic integration of multicolor LEDs or phosphor free multi-junction RGB white LEDs on silicon.

Acknowledgements

The authors thank Dr. S Krishnamoorthy, Mr. F Akyol, and Mr. Y. Zhang for helpful discussions. This work was supported by the Army Research Office (W911NF-13-1-0329) and by the National Science Foundation CAREER award (DMR-1055164).

Figure Captions

Figure 1: (a) Schematic diagram, and (b) energy band diagram with carrier recombination mechanism of polarization doped nanowire UV LEDs.

Figure 2: (a) Schematic diagram, (b) energy band diagram with carrier recombination mechanism, and (c) SEM image of TJ integrated polarization doped nanowire UV LEDs.

Figure 3: Room temperature PL spectrum of TJ integrated polarization doped nanowire UVLED showing emission from AlGaN active region, GaN nanowire base, and InGaN insertion in the TJ. The sharp line at 400 nm is the second harmonic of laser source.

Figure 4: Current-voltage (I-V) characteristic of nanowire UVLED with and without TJ. Inset shows I-V in log scale.

Figure 5: Electroluminescence (EL) measurement under DC current injection from 20 to 100 A/cm$^2$ for UV LEDs without (a) and with (b) TJ. (c) Integrated EL and (d) relative EQE as a function of DC input current density for UV LEDs without (blue circle) and with (red square) TJ. (e) Integrated EL and (f) relative WPE as a function of DC input power density for UV LEDs without (blue circle) and with (red square) TJ.

Figure 6: On wafer power (blue squares) and EQE (red circles) measurements from a 1 mm$^2$ LED device under DC current injection.

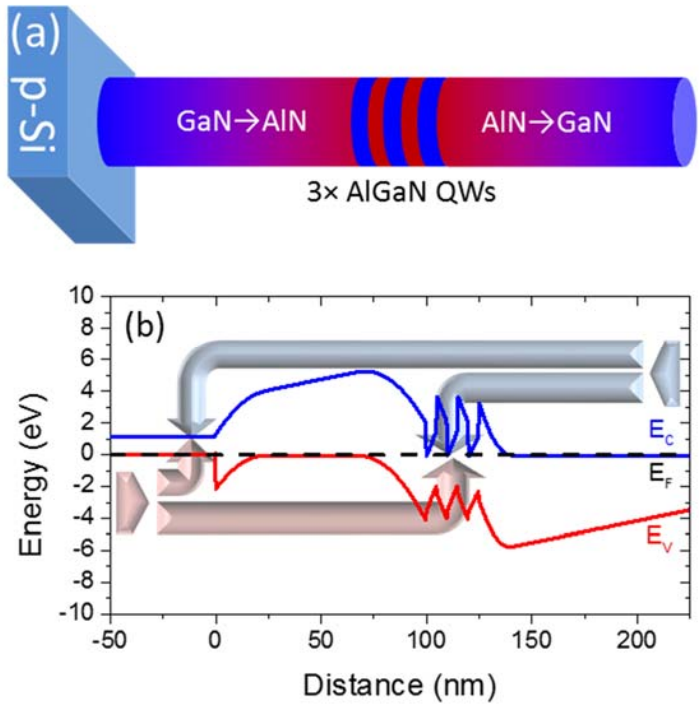

Figure 1 (Single column; Color online)

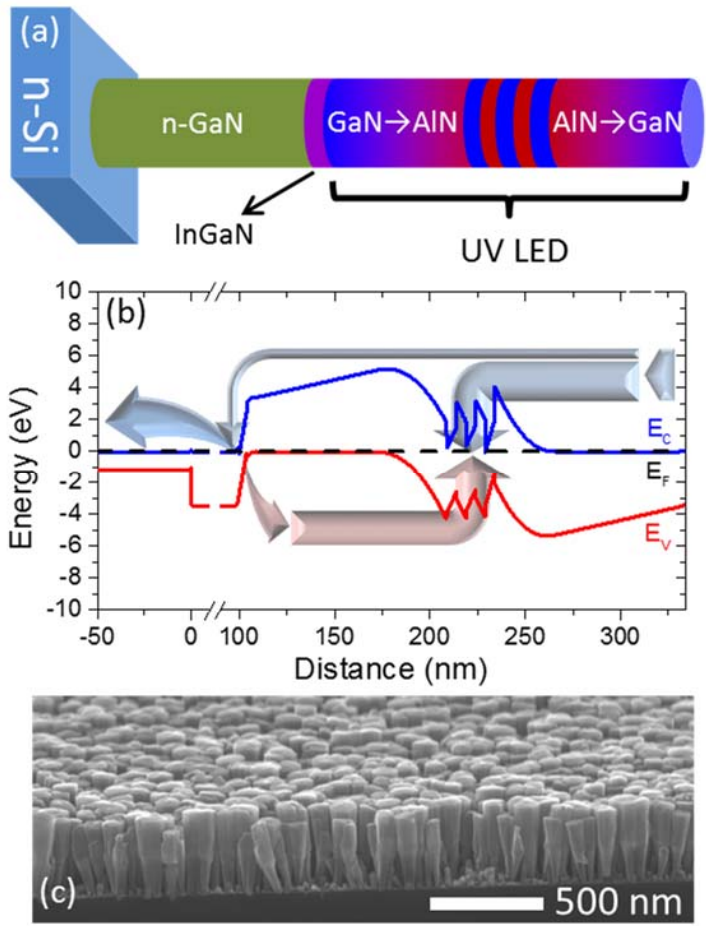

Figure 2 (Single column; Color online)

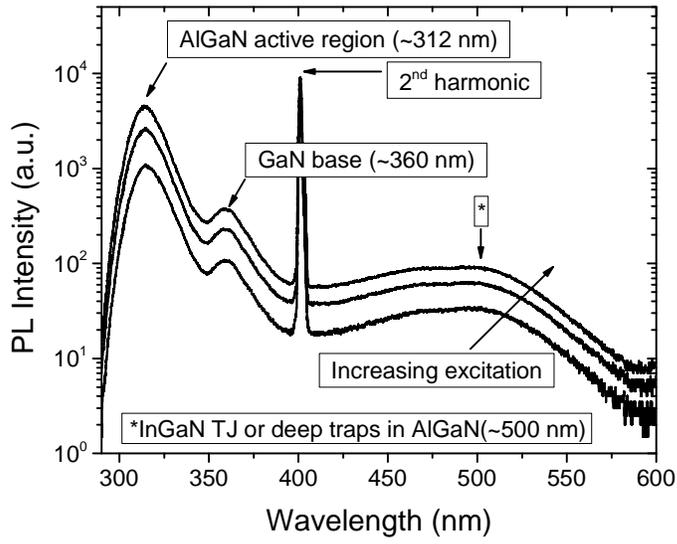

Figure 3 (Single column; Color online)

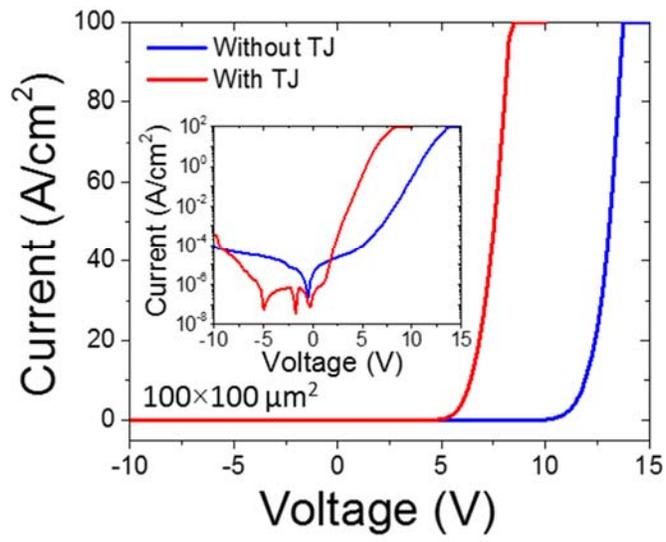

Figure 4 (Single column; Color online)

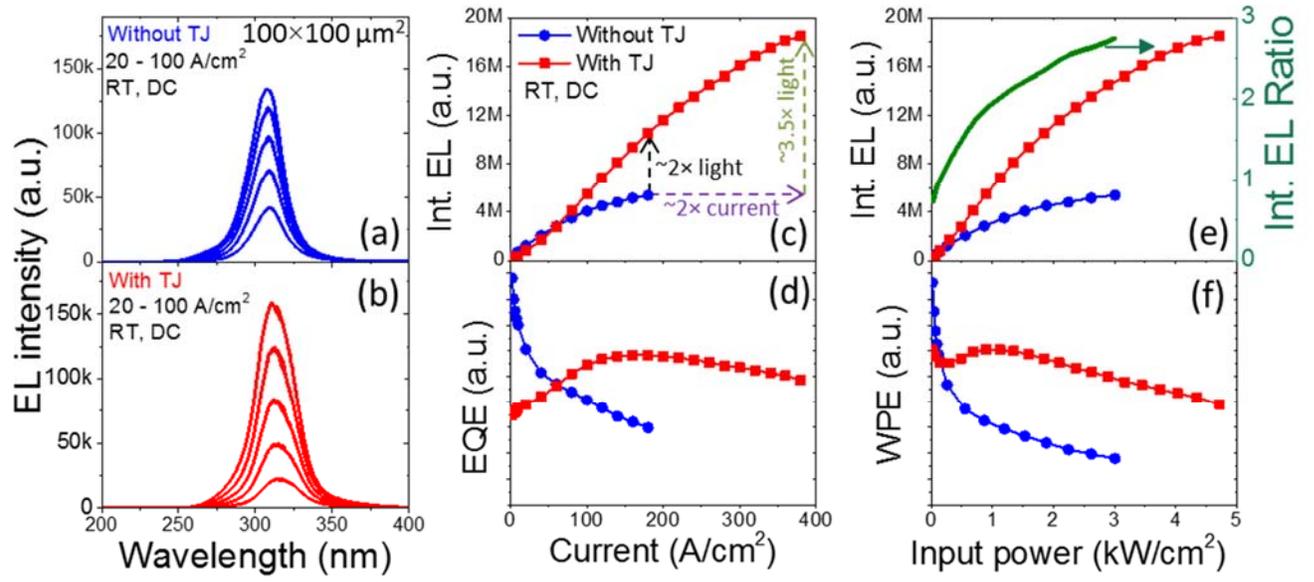

Figure 5 (Double column; Color online)

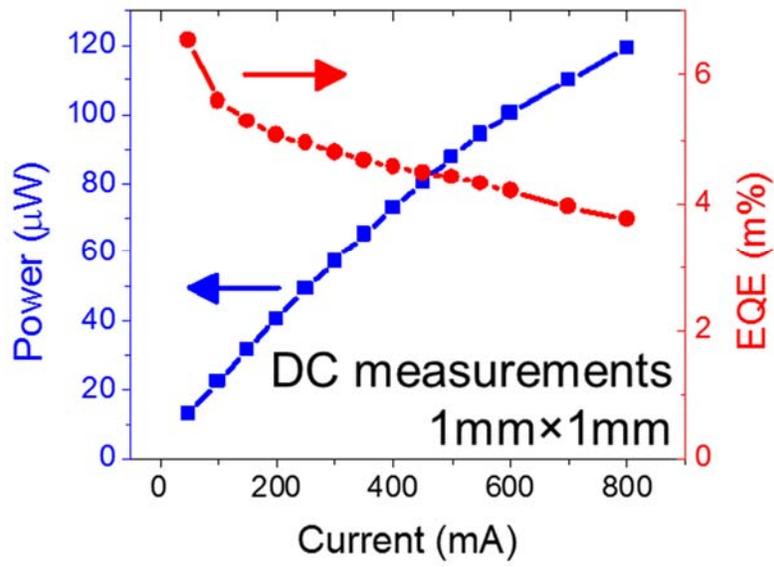

Figure 6 (Single column; Color online)